# Zeolitic imidazolate framework glasses emit white light

Zhencai Li[1*], Zihao Wang[2], Huotian Zhang[3], Xuan Ge[1,4], Ivan Hung[5], Bozhao Yin[6], Fengming Cao[1], Pritam Banerjee[7], Tianzhao, Xu[1], Lars R. Jensen[8], Joerg Jinschek[7], Morten M. Smedskjaer[1], Zhehong Gan[5], Laurent Calvez[9], Guoping Dong[6], Jianbei Qiu[10], Donghong Yu[1], Feng Gao[3], Haomiao Zhu[2*], Yuanzheng Yue[1*]

[1]Department of Chemistry and Bioscience, Aalborg University, DK-9220 Aalborg, Denmark
[2]Xiamen Research Center of Rare Earth Materials, Haixi Institutes, Chinese Academy of Sciences, Xiamen 361021, China
[3]Department of Physics, Chemistry, and Biology (IFM), Linköping University, Linköping 583 30, Sweden
[4]Shanghai Key Laboratory of Materials Laser Processing and Modification, School of Materials Science and Engineering, Shanghai Jiao Tong University, 200240 Shanghai, China
[5]National High Magnetic Field Laboratory, FL 32310, USA
[6]State Key Laboratory of Luminescent Materials and Devices, School of Materials Science and Engineering, South China University of Technology, Guangzhou 510640, China
[7]National Centre for Nano Fabrication and Characterization (DTU Nanolab), Technical University of Denmark (DTU), 2800 Kongens Lyngby, Denmark
[8]Department of Materials and Production, Aalborg University, DK-9220 Aalborg, Denmark
[9]Univ Rennes 1, CNRS, ISCR (Institut des Sciences Chimiques de Rennes) - UMR 6226, F-35000, Rennes, France
[10]Department of Material Science and Engineering, Kunming University of Science and Technology, Kunming 650093, China

**Abstract:**

Zeolitic imidazolate framework (ZIF) glasses represent a newly emerged class of melt-quenched glasses, characterized by their intrinsic nanoporous structure, good processability, and multifunctionalities such as gas separation and energy storage. However, creating photonic functionalities in Zn-based ZIF glasses remains elusive. Here we show a remarkable broadband white light-emitting behavior in a Zn-based ZIF glass, which can be enhanced by annealing. Furthermore, we discovered a sharp red shift upon increasing annealing temperature above the critical temperature of $1.07T_g$, where $T_g$ is the glass transition temperature, for a short duration of 30 min. Finally, we achieved a high absolute internal photoluminescence quantum yield of 12.2% upon annealing of ZIF glass at $1.13T_g$. Based on the optimally annealed ZIF glass, we fabricated a white light-emitting diode (LED) with the luminous efficacy of 4.2 lm/W and high operational stability, retaining 74.1% of its initial luminous efficacy after 180 min of continuous operation. These results not only demonstrate the feasibility of utilizing ZIF glasses in LED applications but also mark a significant advancement in the development of durable, efficient, and multifunctional photonic materials.

**Introduction**

Solid-state white light-emitting diodes (LEDs) have emerged as an attractive solution for next-generation illumination owing to their exceptional energy efficiency, long operational lifespans, and environmental sustainability (1). They are typically fabricated by combining a blue GaN-based LED with either a single yellow phosphor or a combination of red and green phosphors. Thus, phosphors play a crucial role in white LEDs because their performances directly determine the luminous efficacy, color rendering index (CRI), color temperature, reliability, and device lifespan (2). Nowadays, various types of white LED phosphors have been developed, which can be classified into inorganic crystal phosphors (3), organic phosphors (4), perovskite nanocrystals (5), and inorganic glass/ceramic-based phosphors (6, 7). Each phosphor has achieved impressive performance along many dimensions but suffers from distinct limitations, such as low thermal and chemical stabilities. To overcome these, it is important to develop new types of luminescent materials.

Metal-organic frameworks (MOFs) are microporous crystals composed of metal ions coordinated to organic ligands, possessing a large surface area that enables various applications in gas sorption (8-10), gas separation (11), catalysis (12), and ion transportation (13). Some zeolitic imidazolate frameworks (ZIFs) (a subset of MOFs) can be vitrified to glass state via melt-quenching, with their porosity being inherited from the crystal state to a great extent (14, 15). For instance, ZIF-4 ($M(Im)_2$) and ZIF-62 ($M(Im)_{2-x}(BIm)_x$) are good glass formers, where M is the central metal node such as $Zn^{2+}$ or $Co^{2+}$, and Im and BIm are the linkers of imidazole ($C_3H_3N_2^-$) and benzimidazole ($C_7H_5N_2^-$), respectively. ZIF-62 demonstrates ultrahigh glass-forming ability and a wide temperature window of molten state between its liquidus and decomposition temperatures (16, 17). Thus, ZIF-62 can be easily vitrified, e.g., by using the spark plasma sintering (SPS) technique (18). Owing to the $3d^{10}$ close-shell configuration of Zn, Zn-based ZIF crystals have been considered photoluminescently inactive in the visible regions (19). However, given that both Im and BIm show luminescence due to their outstanding optical absorption and π-π* electronic transition in *p*-orbital (20), the Zn-based ZIF-62 crystal is expected to exhibit luminescent properties, which differ from those of its linkers. Its luminescence behaviour depends on metal-ligand interactions, such as metal-to-ligand charge transfer (MLCT) or ligand-to-metal charge transfer (LMCT). Nevertheless, ZIF-62 crystal only exhibits strong ultraviolet light emission, but not visible light emission for white LEDs, because of its too wide band gap (around 3.5 eV).

In this work, we discovered broadband visible light emission in ZIF-62 glasses, which was much stronger than that in their crystalline counterparts. This function was significantly enhanced by annealing at $1.07T_g$ (where $T_g$ is the glass transition temperature) and simultaneously a red shift was induced. We clarified the structural and electronic origin of both the emission enhancement and the red shift. Finally, we succeeded in fabricating ZIF-62 glass-based white LEDs with high absolute photoluminescence quantum yield (PLQY) and good thermal stability.

## Results and Discussion

### Characterizations of ZIF-62 and glasses

Fig. 1A shows the schematic unit cell structure of a ZIF-62 crystal. The synthesized ZIF-62 was used for fabricating melt-quenched ZIF-62 glass (MQG) (21, 22). Then, hot-pressed glass (HPG) was obtained by subjecting the MQG to the SPS process. Both HPG and MQG (in tablet form) were annealed at various temperatures above their $T_g$, noted as $HPG_{xxx\text{-}yy}$ and $MQG_{xxx\text{-}yy}$, where *xxx* denotes the annealing temperature (in K) and *yy* refers to the annealing time (in min), respectively. Fig. 1B shows the preparation procedure of the studied samples. Their characteristics are shown in tables S1 and S2, where their chemical compositions were determined through liquid $^1$H NMR spectroscopy (fig. S1), their crystalline or amorphous structure was identified from the X-ray diffraction (XRD) analysis (Fig. 1D), and $T_g$ and melting point ($T_m$) were determined from the differential scanning calorimetry (DSC) curves in Fig. 1C.

The endotherm (between 440-600 K) in the DSC upscan curve of ZIF-62 crystal (Fig. 1C) is caused by the desolvation of DMF, whereas the endotherm (between 684-719 K) is due to melting (16). Upon hot-pressing (HP) at 60 MPa, the $T_g$ of MQG decreases from 593 K to around 550 K. This decrease was attributed to the pressure-induced weakening of the intermediate-range structural connectivity (18). Upon annealing at 673 K (i.e., 80 K above ambient pressure $T_g$) for 30 min, $T_g$ of HPG increases to around 563 K, implying that the partially distorted structure recovers to a more stable state (fig. S2). The XRD patterns (Fig. 1D) and the high-energy synchrotron XRD (HEXRD) patterns (fig. S3A) confirm the crystalline nature of ZIF-62 and the non-crystalline nature of MQG, HPG, and $HPG_{673-30}$ samples. Further, the XRD patterns (fig. S3B) prove that HPG annealed at various temperatures remains glassy. We also find that the tetrahedral Raman active vibrations in ZIF-62, MQG, HPG, and $HPG_{673-30}$ samples occur at nearly the same frequencies with similar amplitude (Fig. 1E), confirming the integrity of organic linkers in the glasses (16, 23). However, compared to the Zn-N stretching modes of crystalline ZIF-62, those of the three glasses slightly shift to a higher frequency, while the C-N modes shift to a lower frequency, indicating that both annealing and compression make the coordination bonds stable (figs. S4A, B, and D). Interestingly, the three treated glasses exhibit stronger fluorescence than that of crystalline ZIF-62, with $HPG_{673-30}$ featuring the strongest fluorescence. This highlights the significant impact of the subtle tetrahedral rearrangement on the electronic structure of the glasses (fig. S4C), leading to enhanced photoluminescence (PL). We characterize the crystal-to-glass transition of ZIF-62 by in-situ heating transmission electron microscopy (TEM). The morphologies of ZIF-62, MQG, and $MQG_{673-30}$ samples are shown in the first column in figs. S5A, B, and C, respectively. The selected area electron diffraction (SAED) patterns (second column in figs. S5A, B, and C) confirm the ordered structure of ZIF-62 crystal and the disordered structure of MQG and MQG$_{673-30}$, respectively (24). Energy dispersive X-ray spectroscopy (EDS) mapping analysis, as collected from the high-angle annular dark-field imaging scanning TEM (HAADF STEM) images, confirms the homogeneous distribution of Zn, C, and N elements in the samples (the fourth, fifth, sixth columns in figs. S5A, B, and C).

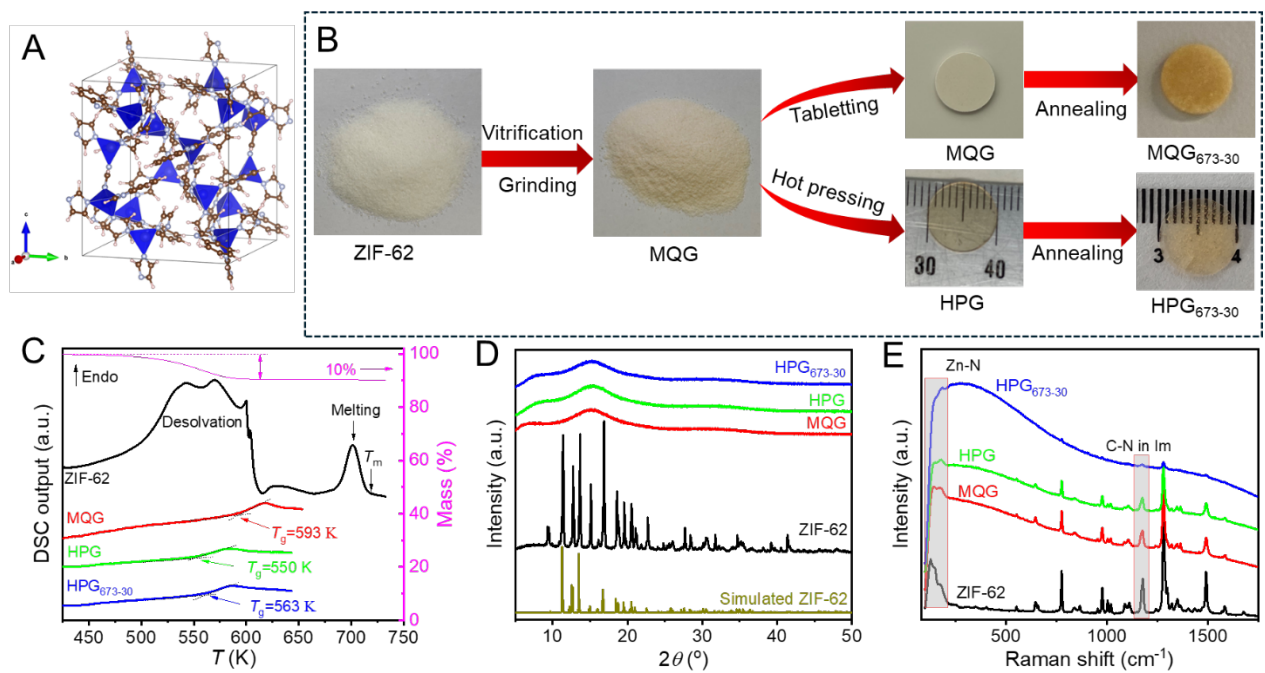

**Fig. 1 Synthesis and characterization of ZIF-62 crystal and glasses.** (A) Unit cell structure of ZIF-62 crystal. (B) Flow chart of melting applied to ZIF-62, hot-pressing (HP) applied to melt-quenched glass (MQG) powder, tableting and annealing applied to MQG, and annealing applied to hot-pressed glass (HPG). (C) Thermogravimetric curve (magenta color) of ZIF-62 crystal and DSC curves of ZIF-62, MQG, HPG, and HPG$_{673-30}$. (D) XRD patterns of simulated ZIF-62, ZIF-62, MQG, HPG, and HPG$_{673-30}$. (E) Raman spectra of ZIF-62, MQG, HPG, and HPG$_{673-30}$.

## Photoluminescence performance of ZIF-62 crystal and glasses

Figs. 2A, B, C, and D show the two-dimensional (excitation-emission) fluorescence spectra of ZIF-62, MQG, HPG, and HPG$_{673-30}$, respectively. ZIF-62 exhibits a narrow emission peak at 300 nm under excitation at the wavelength of 250 nm and extremely weak broad emission peaks under excitation at 320, 370, and 410 nm (fig. S6A). Their emission positions coincide with the edge position of their absorption spectra (fig. S7A) (25). Interestingly, the weak broad emissions are strongly enhanced and red-shifted through vitrification, HP, and annealing at 673 K for 30 min (Figs. 2B-D and figs. S6B-D). However, the narrow peak of native ZIF-62 weakens with vitrification, HP,

and annealing at 673 K for 30 min, which is attributed to the lack of periodic arrangement of Im/BIm rings, suppressing π*-π transitions in BIm rings (Fig. 2E and fig. S8B). The enhancement of the broad emission peaks is assigned to the enhanced ligand-to-metal charge transfer (LMCT, from N to Zn atoms) by the vitrification-, HP-, and annealing-induced structural disorder. The red-shifting of the broad emission peaks in the MQG, HPG, and HPG$_{673-30}$ can be attributed to the red-edge effect (26), leading to a decrease in the gap energy between the π-π* molecular orbitals in Im rings (Fig. 2E and fig. S8B).

However, the narrow emission peaks for ZIF-62 under excitation at 250 nm slightly shift to a shorter wavelength upon vitrification, and HP (figs. S6A). This blue shift could result from the scission and renewal of some Zn-N coordination bonds during vitrification and HP, leading to a distortion of tetrahedral units in MQG and HPG (27). In contrast to the blue shift, red shifts of the broad emission peaks are found in HPG annealed at 673 K for 30 min under excitation at 320, 370 and 410 nm (figs. S6B-D). In particular, such red shifts appear at a much higher wavelength compared to MQG and HPG samples annealed at lower temperatures (figs. S9-10).

The optical images in Fig. 2F illustrate the emission colors of ZIF-62, HPG, and HPG$_{673-30}$ under excitation of both natural light and 365 nm laser. Upon vitrification, HP, and annealing at 673 K for 30 min, the broad emission of ZIF-62 is significantly enhanced, accompanied by a color shift from blue to cold-white. These results indicate that the HPG$_{673-30}$ is a highly promising candidate for application in white LEDs. The chromaticity coordinates of ZIF-62 (0.210, 0.216), MQG (0.195, 0.179), HPG (0.205, 0.202), and HPG$_{673-30}$ (0.256, 0.295) under 370 nm excitation are marked in the CIE 1931 color spaces (Fig. 2G). The color point of HPG$_{673-30}$ lies in the cold-white lighting region, whereas those of ZIF-62, MQG, and HPG are located in the blue lighting region. The chromaticity coordinates of ZIF-62, MQG, HPG, and HPG$_{673-30}$ under 250, 320, and 410 nm excitation are shown in fig. S11. We find that ZIF-62, MQG, HPG, and HPG$_{673-30}$ emit much weaker cold-white light under 250 nm excitation compared to other excitation wavelengths (fig. S6A). Therefore, HPG$_{673-30}$ is regarded as an excellent candidate for cold-white light emitting under 370 nm excitation.

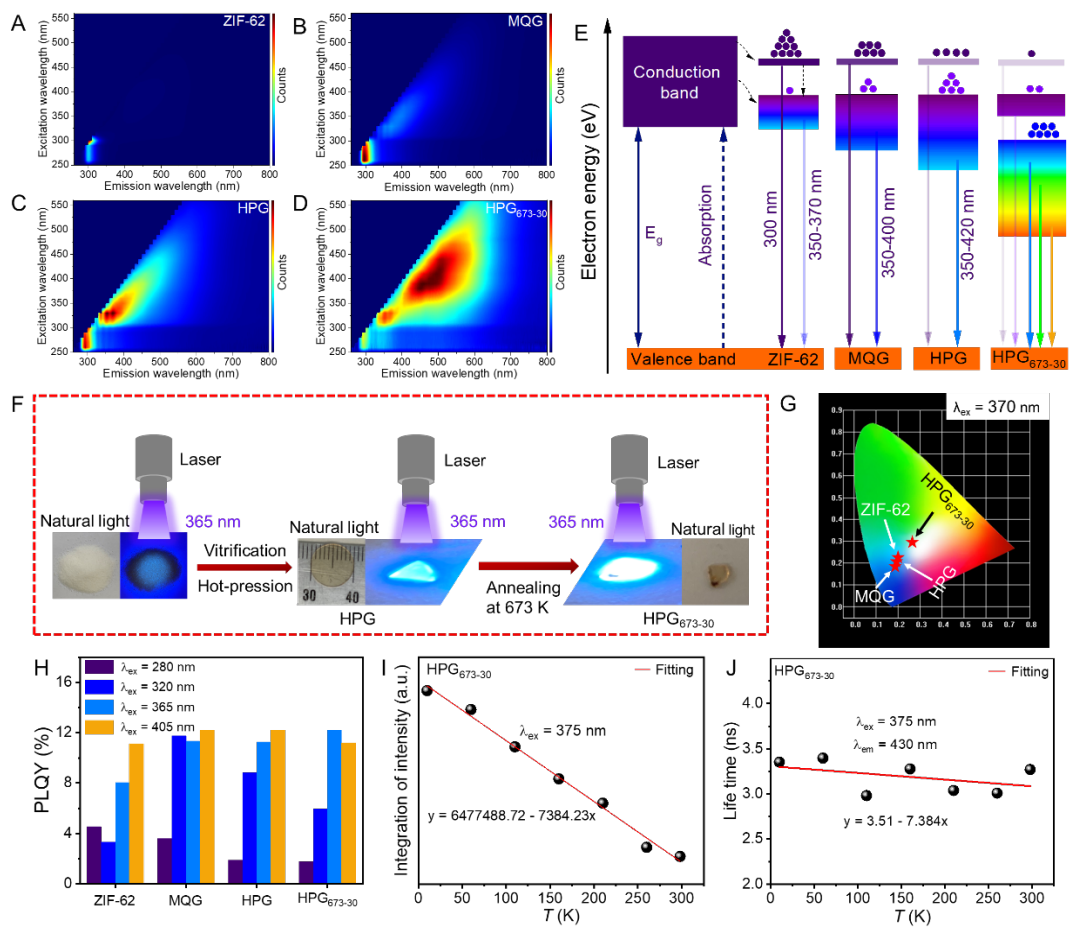

**Fig. 2 Photoluminescence (PL) performance of ZIF-62 crystal and glasses.** (A-D) Two-dimensional fluorescence (Excitation/Emission) spectra of (A) ZIF-62, (B) MQG, (C) HPG, and (D) HPG$_{673-30}$. (E) Proposed PL mechanism of ZIF-62, MQG, HPG, and HPG$_{673-30}$. (F) Optical photos of ZIF-62, HPG, and HPG$_{673-30}$ under natural light and 365 nm laser. (G) The Commission Internationale de l'Éclairage (CIE) chromaticity coordinates of PL spectra of ZIF-62, MQG, HPG, and HPG$_{673-30}$ under 370 nm excitation. (H) Absolute PLQY of ZIF-62, MQG, HPG, and HPG$_{673-30}$ under excitation at 280, 320, 365, and 405 nm, respectively. (I and J) Temperature-dependent integral PL (m) and lifetime (n) of HPG$_{673-30}$ under excitation of 375 nm. The curves represent linear fits as a function of temperature.

Fig. 2H and table S3 show the absolute internal PLQY values for ZIF-62, MQG, HPG, and HPG$_{673-30}$ under excitation at 280, 320, 365, and 405 nm, respectively. The internal PLQY of ZIF-62 under excitation at 280 nm sharply decreases from 4.52 to 1.75% upon vitrification, hot-pressing, and annealing at 673 K for 30 min. In contrast, the internal PLQY of ZIF-62 crystal under excitation at 320 nm first increases from 3.32 to 11.74% upon vitrification and then decreases to 5.94% by hot-pressing and annealing at 673 K for 30 min. Evidently, it is enhanced from 8.04 to 12.19% upon vitrification, hot-pressing, and annealing at 673 K for 30 min for ZIF-62 under excitation at 365 nm. However, it remains nearly unchanged (around 12 %) in ZIF-62, MQG, HPG, and HPG$_{673\text{-}30}$ under excitation at 405 nm. Similarly, the external PLQY under 365 nm excitation (table S4) is enhanced in ZIF-62 upon vitrification, HP, and annealing at 673 K for 30 min. The fluorescence lifetime for ZIF-62 at 430 nm under excitation at 375 nm (fig. S12) is extended by vitrification, HP, and annealing at 673 K for 30 min.

Additionally, we find that the PL intensity of HPG$_{673\text{-}30}$ decreases with increasing measurement temperature (fig. S13), due to the increased possibility of non-radiative transition. By integrating the PL over the wavelength, we find that the integrated value linearly decreases with temperature (Fig. 2I). Similarly, the fluorescence lifetime of emission at 430 nm for HPG$_{673\text{-}30}$ linearly decreases with increasing temperature (Fig. 2J). This temperature-dependent PL and lifetime behavior indicate that MQG$_{673\text{-}30}$ (fig. S14) also exhibits excellent thermal stability in the typical operating temperature window of LEDs (around 400 K).

**LED performance of ZIF-62 glasses**

Based on the above-reported promising results, we fabricated six LEDs (LED-1, -2, -3, -4, -5, and -6) based on ZIF-62 glasses (see Methods Section) and evaluated their performances. Table S5 presents the photoelectric parameters of the three LEDs as well as a commercial white LED (CWLED), including correlated color temperature (CCT), color rendering index (CRI, denoted as $R_a$), and luminous efficacy. Figs. 3A-B show the PL spectra of LED-1 (combining bulk HPG and a 365 nm ultraviolet chip, an LED as an excitator for exciting the LED) and LED-2 (integrating bulk HPG$_{673\text{-}30}$ and a 365 nm ultraviolet chip), respectively, under a driving current of 60 mA. The chromaticity coordinates of LED-1 (0.231, 0.254) and LED-2 (0.254, 0.300) appear in the blue and cold-white lighting regions in CIE 1931 (Fig. 3F). Compared with the CWLED utilizing a single

YAG: $Ce^{3+}$ phosphor (28), LED-1 and LED-2 exhibit the advantage of higher $R_a$. However, their CCTs are significantly higher than those of CWLED, leading to a pronounced cold tone of the emitting light owing to the weak red-light component. Fig. 3C shows the PL spectrum of LED-3, which integrates bulk $HPG_{673-30}$, a 365 nm ultraviolet chip, and the commercial red phosphor ($CaAlSiN_3$: $Eu^{2+}$), under a driving current of 100 mA. By adding the commercial red phosphor in LED-3, the CCT decreases to 8448 K and $R_a$ increases to 91.8. This indicates that a slightly warm white LED-3 is achieved under the chromaticity coordinate (0.294, 0.293), as shown in CIE 1931 (Fig. 3F), but exhibits a much smaller luminous efficacy (3.15 lm/W) than CWLED. The photographs of the lighted LED-1, LED-2, and LED-3 are shown in insets of Figs. 3A-C, respectively. The zoom-in optical image of the lighted LED-2 is shown in Fig. 3D, while those of the as-fabricated LED-1, LED-2, and LED-3 are shown in fig. S15. The PL spectra and photoelectric parameters of LED-4 (combining 400 nm chip and HPG), LED-5 (combining 400 nm chip and $HPG_{673-30}$), and LED-6 (combining 400 nm chip, $HPG_{673-30}$, and commercial red phosphor) are shown in fig. S16 and table S6, respectively. Optical images of the lighted LED-4, LED-5, and LED-6 are shown in fig. S17.

To evaluate the electric current-dependent LED performance, the PL spectra and luminous efficacies of LED-1, LED-2, LED-4, and LED-5 are measured under various driving currents from 10 to 100 mA, respectively. The luminous efficacy of the fabricated LEDs gradually decreases with increasing driving current (figs. S18-21) due to the decrease in external quantum efficiency (29). Fig. 3e demonstrates the decay of an initial luminous efficacy ($L_0$) of 4.2 lm/W for LED-2 at the constant driving current of 60 mA with continuous operation time. As also shown in Fig. 3e, the real working luminous efficacy ($L_1$) drops to 3.1 lm/W after 180 min operation, remaining at 74.1% of $L_0$. More importantly, the real working luminous efficacy of LED-2 decreases slowly when extending the operating time, implying its excellent operational stability. This performance indicates the potential use of annealed ZIF-62 glass in low-power LEDs, such as electronic displays, night lighting, indicator lighting, and horticultural lighting.

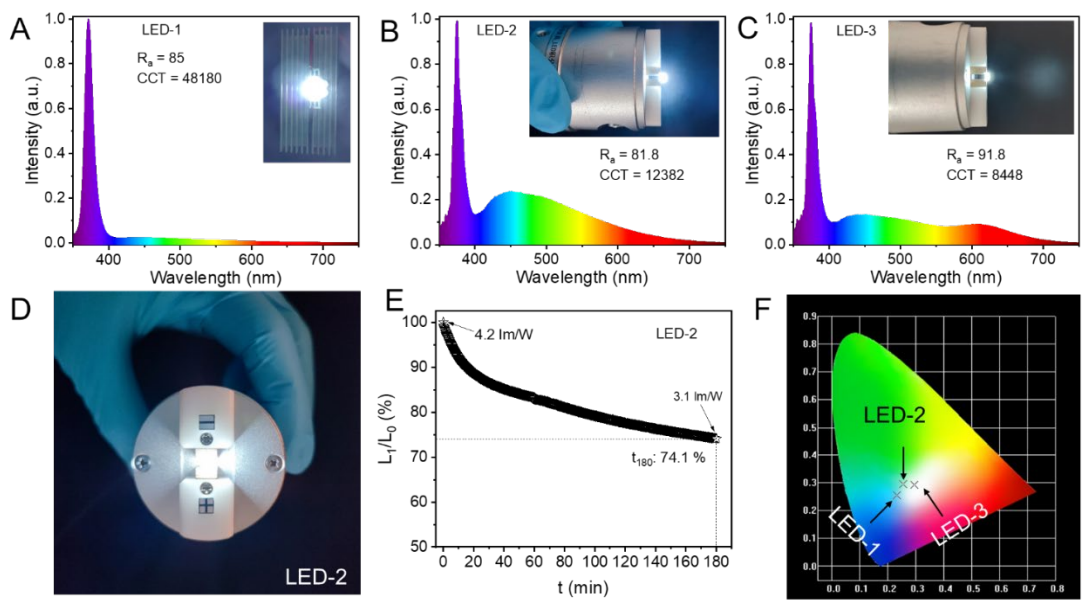

**Fig. 3 Device performance of fabricated white LEDs based on ZIF-62 glasses.** (A-B) PL spectra of LED-1 (A) and LED-2 (B) under 60 mV driving current, with photographs of the lighted LEDs shown in the insets. (C) PL spectrum of LED-3 (made by adding the commercial red phosphor ($CaAlSiN_3$: $Eu^{2+}$)) under 100 mV driving current. (D) Photograph of the LED-2. (E) Operational stability of LED-2 under continuous operation at a driving current of 60 mA. (F) Chromaticity coordinates of the three as-fabricated LEDs marked in CIE 1931 color spaces.

## Structure analyses of ZIF-62 crystal and glasses

To elucidate the difference in the electronic structure variations underlying the changes in PL, the atomic-scale (short-range) structure of ZIF-62 crystal and the three glasses were first characterized by solid-state $^{67}Zn$, $^{15}N$, and $^{13}C$ NMR spectroscopy. Compared to ZIF-62, the $^{67}Zn$ NMR spectra (Fig. 4A and fig. S21A) of MQG, HPG, and $HPG_{673-30}$ have broadened asymmetric line shapes with low-frequency tails, and the resonance peak shifts to a lower frequency. The broadening and characteristic line shape is attributed to a continuous distribution of $C_Q$ due to structural disorder in the glassy state, verifying the short-range disorder of [N-Zn-N] tetrahedra in ZIF-62 glasses (30). However, there is no detectable difference in the $^{67}Zn$ NMR spectra among the three ZIF-62 glasses, indicating that chemical integrity was retained upon HP and annealing. Compared to the resonance peak (around 204 ppm) for N (in Im) in the $^{15}N$ NMR spectrum of crystalline ZIF-62 (Fig. 4B and fig. S22), those of the three glasses are broader and shifted to a higher chemical shift of 205 ppm. This reveals the de-shielding effects on electron clouds around nitrogen atoms, resulting in their increased chemical shifts (31), indicating electron (charge) transfer from imidazole ligands to zinc metal. This transfer is corroborated by a slight decrease in chemical shifts for Zn nuclei of MQG, HPG, and $HPG_{673-33}$ relative to that of ZIF-62 crystal in the data above, suggesting that the density of electron clouds around Zn nuclei is enhanced in the former, likely again due to the electron (charge) transfer from ligands to metal. Therefore, the LMCT effects are clearly demonstrated by both $^{67}$Zn and $^{15}$N NMR data, supporting the red shift of the PL spectra of the three treated ZIF-62 glasses. Consequently, the π-electron delocalization is induced, and thereby the conjugation effect within Im and BIm rings is enhanced, resulting in a red shift in PL of the annealed glass (32-34). In addition, another weak resonance peak of N (in BIm) at 190 ppm is broader in the three ZIF-62 glasses, without a detectable shift in the resonance signal (35), further indicating that LMCT occurs in both Im and BIm moieties.

Similar to the $^{67}$Zn and $^{15}$N NMR spectra, two populations of $^{13}$C are found in the $^{13}$C CPMAS NMR spectra of ZIF-62, MQG, HPG, and $HPG_{673-30}$ (Fig. 4C), i.e., C-nuclei in BIm and Im (insets of Fig. 4C) of the native ZIF-62 crystal and the three treated glasses (MQG, HPG, and $HPG_{673-33}$). The most pronounced decreased chemical shifts are ascribed to the carbon nuclei from the imidazole unit (C-8) for the three treated ZIF-62 glasses compared with that for the native ZIF-62 crystal. Such lower chemical shifts correspond to increased electron density, implying possible charge transfer from N to C nuclei, but the relatively small change implies that LMCT is dominating. The difference in $^{13}$C signals indicates structural distortion of Im and BIm rings upon vitrification, hot-pressing, and annealing.

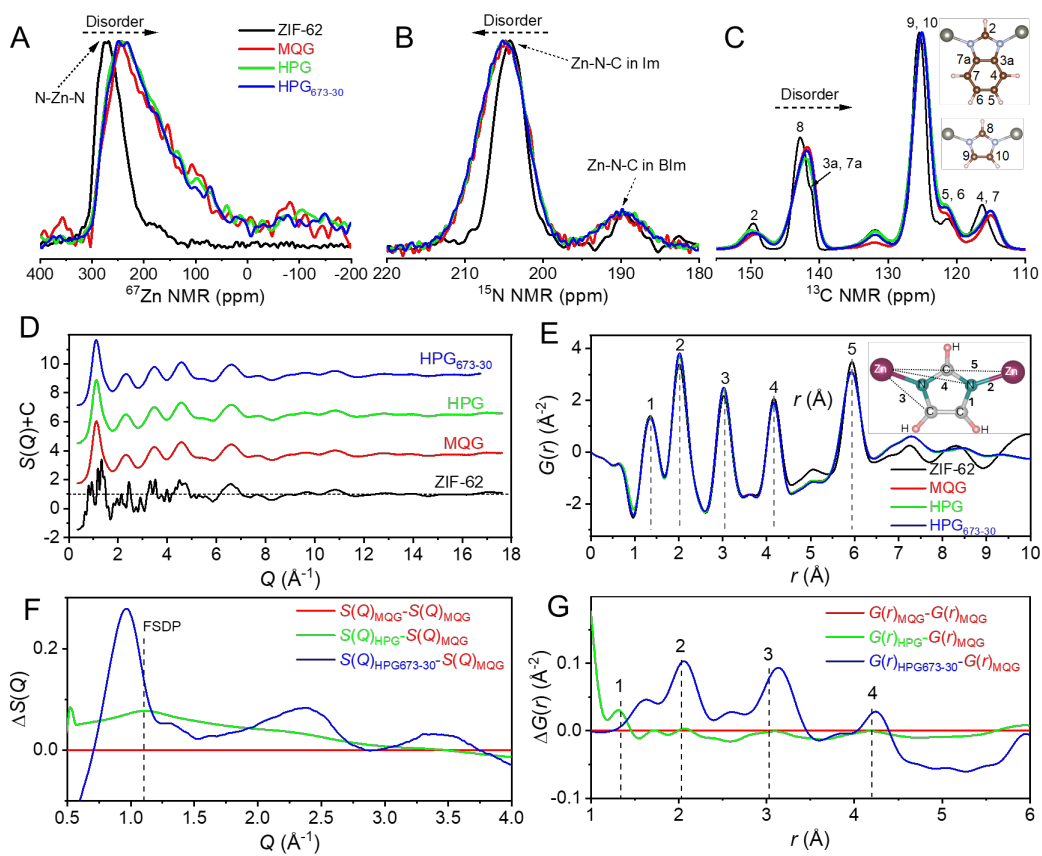

**Fig. 4 Structure analyses of ZIF-62 and glasses.** (A-C) $^{67}$Zn (A), $^{15}$N (B), and $^{13}$C (C) MAS NMR spectra of ZIF-62, MQG, HPG, and HPG$_{673-30}$. (D) Faber-Ziman structure factors $S(Q)$ of ZIF-62, MQG, HPG, and HPG$_{673-30}$. (E) Reduced pair correlation functions $G(r)$ of ZIF-62, MQG, HPG, and HPG$_{673-30}$. (F) The differences in $S(Q)$ ($\Delta S(Q)$) of $S(Q)_{MQG}$-$S(Q)_{MQG}$, $S(Q)_{HPG}$-$S(Q)_{MQG}$, and $S(Q)_{HPG673-30}$-$S(Q)_{MQG}$. (G) The differences in $G(r)$ ($\Delta G(r)$) of $G(r)_{MQG}$-$G(r)_{MQG}$, $G(r)_{HPG}$-$G(r)_{MQG}$, and $G(r)_{HPG673-30}$-$G(r)_{MQG}$, respectively.

To further probe the structural evolution of ZIF-62 caused by vitrification, hot-pressing, and annealing, high-energy synchrotron X-ray diffraction (HEXRD) analysis is performed. The glass nature of MQG, HPG, and HPG$_{673-30}$ is verified by the lack of sharp Bragg peaks in the reciprocal-space $S(Q)$ curves (Fig. 4D). Fig. 4f shows that the first sharp diffraction peak (FSDP) in MQG is enhanced by HP and annealing both at 673 K for 30 min and at other temperatures (fig. S23), implying an increase in the connectivity of [N-Zn-N] tetrahedra, i.e., a higher degree of medium-range order in HPG and annealed HPG (36). The real-space $G(r)$ data (Fig. 4E) of all the samples are obtained by applying the Fourier transformation to the $S(Q)$ data. The atom-atom pair correlations of C-C(N) (1.36 Å, 1), Zn-N (2.02 Å, 2) and (4.17 Å, 4), Zn-C (3.03 Å, 3), and Zn-Zn (5.94 Å, 5) are marked in the $G(r)$ curves. The Zn-Zn correlation in the glasses is notably weaker than that in the crystalline counterpart, confirming that the glasses possess a higher distortion of the intermediate-range structure of Zn-linker-Zn. However, as shown in $\Delta G(r)$ curves (Fig. 4G), the short-range correlations of C-C(N) (1.36 Å), Zn-N (2.02 Å), Zn-C (3.03 Å) and Zn-N (4.17 Å) are visually strengthened upon annealing (fig. S24), implying that the coordination Zn-N and covalent C-C(N) bonds get stronger.

**Mechanism of photoluminescence in ZIF-62 crystal and glasses**

To understand the interplay between electronic structures and PL of the studied samples, X-ray photoelectron spectroscopy (XPS) measurements are conducted for ZIF-62, MQG, and MQG$_{673-30}$. As seen in fig. S25, the peaks in Zn 2p, N 1s, and C 1s spectra of the ZIF-62 crystal become slightly narrower upon vitrification, while they get further narrowed through annealing. The narrowing of the XPS peaks means a more uniform electron distribution around the Zn, C, and N atoms since the short- and intermediate-range structures become more ordered via annealing. This is confirmed by the enhancement of the short-range correlations of C-C(N) and Zn-N(C) in the $\Delta G(r)$ curves (Fig. 4G).

In addition, the electronic structure of the ZIF-62 is calculated using periodic density functional theory (PDFT) to gain insight into the PL mechanism. As shown in Fig. 5D, we observe a flat band electronic structure with a band gap of 3.58 eV. Based on the analysis of the total/partial density of states (DOS) (Fig. 5C), the electronic density is mainly determined by the $p$-orbitals of C and N atoms. Furthermore, the highest occupied molecular orbital (HOMO) and the lowest unoccupied molecular orbital (LUMO) are dominated by the $p$-orbital of C and N in the Im and BIm rings (fig. S26). This indicates that the PL of ZIF-62 crystal is primarily attributed to the $\pi$-$\pi^*$ interaction within the rings. The electrostatic potential map and electronic structure of Im (Fig. 5D) show that 'pyridine-like' N contributes one π electron in its $p$-orbital, forming its aromaticity, while its lone-pair electrons are in a $sp^2$-orbital without being involved in its molecular conjugation. The other 'pyrrole-like' N atom contributes its lone-pair electrons to the conjugated aromaticity. Such different chemical environments of N-nuclei must result in different energy barriers for the "free-rotation" of -N-Zn-N-Zn- bonds. This is indeed proven by the slight splitting of both Zn- and N-peaks in their $^{67}$Zn and $^{15}$N NMR spectra, respectively, as shown in Figs. 4a and b. Such Zn-N ('pyrrole-like') coordination can influence the π electron distribution in the *p*-orbital of N, thereby affecting the PL in crystalline ZIF-62.

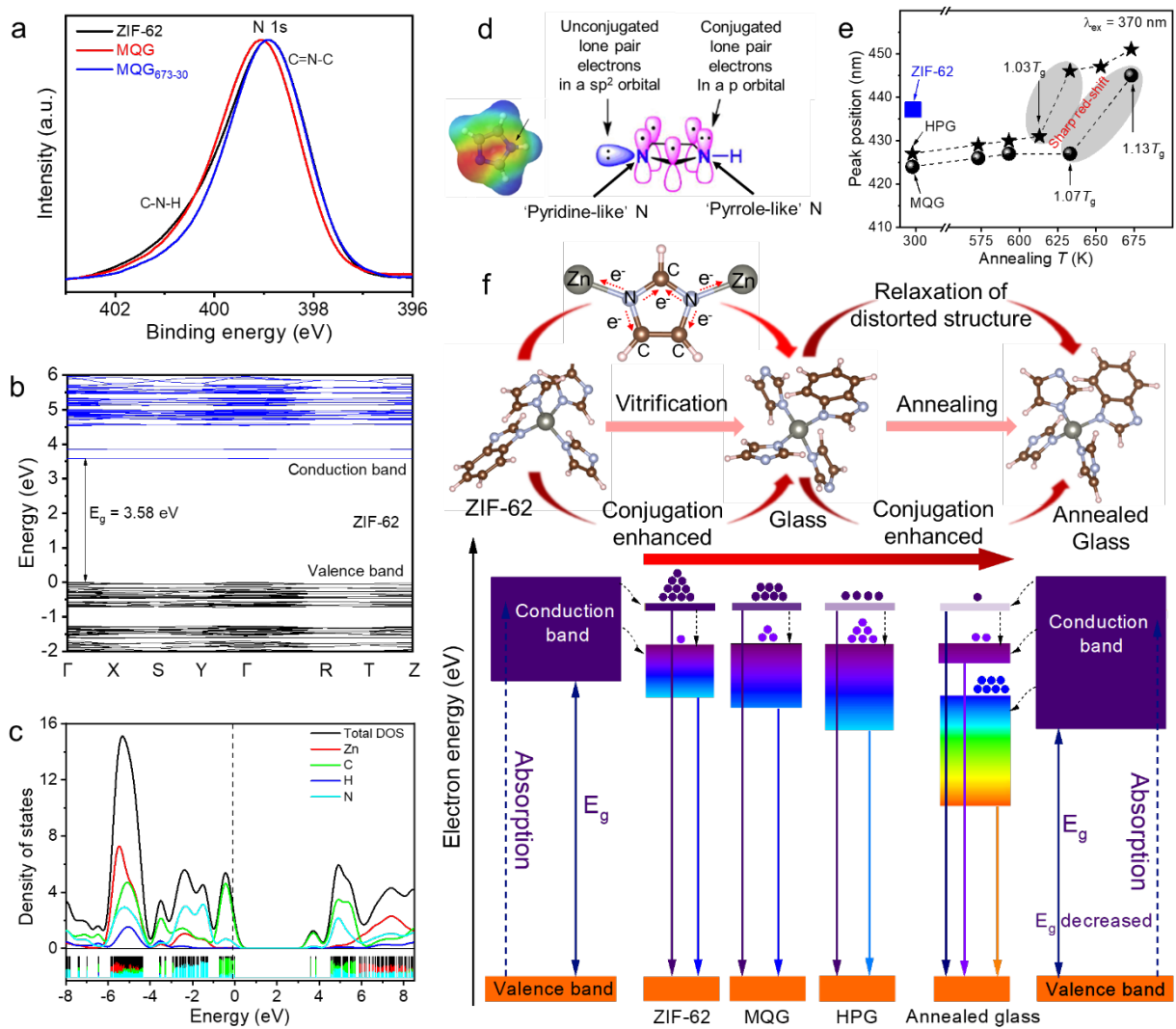

**Fig. 5 Electrical structure and mechanism of PL in ZIF-62 crystal and glasses.** (a) Valence-band XPS N 1s spectra in ZIF-62, HPG, and HPG$_{673-30}$. (b) Band structure and (c) corresponding total/partial density of states (TDOS/PDOS) of ZIF-62. (d) Electrostatic potential map and electronic structure of Im. (e) PL peak position as a function of annealing temperature. (f) Proposed PL mechanism of ZIF-62 upon vitrification, HP, and annealing.

The weak Zn-N ('pyrrole-like') bond involves delocalizing the π-lone pair electrons of N to Zn, therefore increasing its conjugation area. On the other hand, the strong Zn-N ('pyridine-like') bond contributes its conjugation area, but only one π-electron to slightly increase the conjugation area.

Therefore, the charge transfer from N to both Zn and C nuclei occurs to a different extent, causing the increased conjugation area of the entire tetrahedral units in the studied ZIFs (see left upper panel in Fig. 5f). Upon hot-pressing, the weak Zn-N ('pyrrole-like') bond with lower binding energy could be slightly strengthened, accompanied by a reduced intermediate-range structural hindrance and a decreased $T_g$ (18). The pressing-induced strengthening of the Zn-N ('pyrrole-like') bond leads to a delocalization of π electrons, i.e., an increased conjugation area in the *p*-orbital. Consequently, band-gap energy ($E_g$) between the valence band and the conduction band decreases upon hot-pressing, resulting in a slight red shift in the PL of HPG (Fig. 5e).

Fig. 5e shows the dependence of the peak position of PL (under excitation at 370 nm) on the annealing temperature ($T_a$) for 30 min duration in MQG and HPG samples. The annealing temperature-dependent PL spectra of MQG and HPG, excited from 360 to 410 nm, are shown in figs. S27 and 28. The PL peak position in MQG only slightly increases with increasing $T_a$ and then sharply increases above a critical $T_a$, i.e., $1.07T_g$ (633 K). In contrast, in the case of HPG, the PL peak position sharply increases with $T_a$ above $1.03T_g$ (613 K). That is, the sharp red shift of PL in MQG occurs at a $T_a$ that is 20 K higher than that in HPG, implying that HPG undergoes a higher degree of distortion of the intermediate-range structure (around 0.5-2 nm) compared with MQG. The activation energy for the relaxation of the distorted structure towards a less distorted structure is lowered by hot-pressing, and hence, a lower $T_a$ is needed for relaxing HPG compared to MQG.

The existence of the critical $T_a$ for causing the sharp red shift in PL could be interpreted in terms of relaxation kinetics. The critical $T_a$ of 633 K for MQG corresponds to the viscosity of $10^{8.5}$ Pa·s, as determined by extrapolating the literature viscosity-temperature data using the Mauro–Yue–Ellison–Gupta–Allan (MYEGA) model (16, 37). At this viscosity, the distorted intermediate-range structure has a high probability of fulfilling the kinetic (diffusional) condition for rearranging itself to a less distorted state during annealing for a certain period. Such structural rearrangement should be concomitant with the translational and rotational motions of the organic linkers towards energetically favorable positions. During such structural rearrangement, a small amount of undercoordinated Zn and N atoms in free organic linkers should be re-coordinated (38-41), forming a stable coordination bond with a slightly higher binding energy. Consequently, some Zn-N ('pyrrole-like') bonds become stronger, thereby increasing the conjugation domains of π electrons in the *p*-orbital (see the right-upper panel in Fig. 5f) and decreasing $E_g$ between the valence band and the conduction band (see the lower panel in Fig. 5f). This ultimately induces a red shift in PL, thereby exhibiting a slightly yellow shade in appearance (see Fig. 1b). We note that the strengthening of the Zn-N bond can be inferred by the enhancement of the correlations of the Zn-N bonds as demonstrated by the $\Delta G(r)$ curves of HPG in Fig. 4G and fig. S24. For the above reasons, the extent of the red shift of PL increases with $T_a$ above $1.07T_g$. To further improve the performance of the MOF glass-based LEDs, the optimal $T_a$ still needs to be identified, which can maximize their PL red shift and quantum yield. However, it is important to note that the upper limit of $T_a$ should not exceed the decomposition temperature of MOF glasses.

**Conclusions**

We synthesized ZIF-62 crystals via the solvothermal approach and prepared ZIF-62 glasses through melt-quenching, hot-pressing, and annealing. We discovered the ultrastrong narrowband ultraviolet emission and the weak broadband visible light emission from the ZIF-62 crystal. More importantly, the broadband visible light emission was enhanced by means of vitrification, hot-pressing, and annealing, while a significant red shift was induced by annealing at a temperature above $1.07T_g$. These phenomena were attributed to the band gap decrease caused by an extension of the conjugation area from the aromatic moieties of Im and BIm rings to the weak Zn-N ('pyrrole-like') bond. Such enhancement of the Zn-N interactions was verified by high-energy synchrotron X-ray diffraction measurements.

Based on these findings, we fabricated broadband white LEDs based on the ZIF-62 glass annealed at $1.13T_g$ (673 K) for 30 min. This annealed glass exhibits a high internal photoluminescence quantum yield of 12.2%. This optimized LED exhibited the luminous efficacy of 4.2 lm/W and excellent operational stability, remaining 74.1% of its original luminous efficacy after 180 min of operation. This work demonstrated that properly processed ZIF glasses are promising candidates for developing high-performance solid-state white LEDs. Future advancements, such as doping with rare-earth or transition metal ions and optimizing contact engineering, could further enhance the performance of ZIF glass LED, making them competitive with mature LED technologies.

## MATERIALS AND METHODS

### Materials

Zinc nitrate hexahydrate ($Zn(NO_3)_2 \cdot 6H_2O$, ≥99%), benzimidazole (BIm, 98%), and imidazole (Im, ≥99%) were purchased from Sigma-Aldrich. N, N-dimethylformamide (DMF, ≥ 99.5 %) and dichloromethane (DCM, ≥99.5 %, amylene as stabilizer) were supplied by VWR Chemicals.

### Synthesis of ZIF-62 crystal

ZIF-62 crystal was synthesized by using the same methodology as given in ref (42). 38.02 g Im and 12.79 g BIm were dissolved in 480 mL DMF and mixed under stirring for 5 min. Then, 19.93 g $Zn(NO_3)_2 \cdot 6H_2O$ was added to the mixed solution. Subsequently, the mixture was stirred until complete dissolution. The resulting synthetic Zn:Im:BIm molar ratio was approximately 3:25:5. The solution was transferred into a 500 mL glass jar, which was then sealed tightly and heated to and kept at 403 K for 72 hr in an oven. The white as-synthesized sediment was separated from the suspension by using centrifugation and thoroughly washed 2 times with DMF and 1 time with DCM. Finally, the obtained ZIF-62 crystal powder was dried at 403 K for 24 hours in an oven.

### Preparation of ZIF-62 glass

The crystalline ZIF-62 powder was melted at 723 K for 5 min under argon in a tube furnace and then quenched naturally to room temperature to obtain ZIF-62 glass. Such melt-quenched glass (MQG) was crushed and ground for obtaining fine powder to be used as starting powder for spark plasma sinering (SPS) densification.

**Preparation of transparent hot-pressed glass by SPS**

The sintering process of transparent hot-pressed glass (HPG) was conducted under vacuum on an HP-D 10 SPS instrument (FCT System GmbH). Around 0.35 g ZIF-62 glassy powder was filled into graphite dies (10 mm in diameter), and a K-type thermocouple was inserted centrally in a hole in a wall of the die. The process chamber was evacuated, and the temperature was increased at a rate of 50 K min$^{-1}$ by resistive heat generated from passing a pulsed direct current (max. 0.35 kA, 4.65 V) through the graphite part. After 2 min dwelling, the sample was cooled at about 50 K min$^{-1}$. In this work, transparent HPGs were sintered at 593 K under 63.67 MPa. The sample named HPG is used in contrast to MQG. It should be mentioned that the temperature difference between the glass sample and the temperature sensor is only 5 K.

**Calorimetric analysis**

The DSC measurement of all the samples was performed under argon using a Netzsch STA 449 F1 instrument. The powder or thin bulk samples were placed in a platinum crucible situated on a sample holder of the DSC at room temperature. For the first scan, the samples were kept for 5 min at an initial temperature of 313 K, heated to the final temperature (723 K) at 10 K min$^{-1}$ and kept for 5 min, and then cooled down to 473 K by 10 K min$^{-1}$, thus forming MQG. After natural cooling down to room temperature, the second scan was conducted using the same procedure as the first scan. To determine the $C_p$ of the samples, both the baseline (blank) and the reference sample (sapphire) were measured.

**Powder X-ray diffraction (XRD) measurements**

Room temperature XRD measurements were performed to identify the crystallization phase by a Malvern-Panalytical diffractometer, operated at 45 kV and 40 mA, with Cu-Kα (λ = 1.5406 Å) radiation during the 2$\theta$ range of 5–50° with a step size of 0.013°.

**Transmission electron microscopy (TEM) measurements**

**In-situ heating in TEM**

The ZIF-62 powder was dispersed in ethanol, ultrasonicated, and drop casted in amorphous-carbon-coated Au TEM grid. The in-situ heating experiment was done using Gatan INCONEL holder in a ThermoFisher Titan microscope operated at 300 kV. The sample was heated from room temperature to 450 °C at a heating rate of 5 °C per second and soaked for 1 h. The TEM images and the SAED data, at both room temperature and elevated temperature, were acquired at a dose rate of 1e$^-$ Å$^{-2}$s$^{-1}$ and with an exposure time of 1 s, respectively.

**HAADF imaging and EDX analysis**

The ZIF-62 and MQG powders were dispersed in ethanol, ultrasonicated, and drop casted in amorphous-carbon-coated Cu TEM grid. The HAADF STEM images and EDS elemental mapping were acquired using ThermoFisherScientific Spectra Ultra microscope operated at 300 kV, with a probe convergence angle of 15 mrad and a camera length of 110 mm. The 4kx4k HAADF STEM images were acquired with a dwell time of 2 μs. Elemental EDS mapping was done using an Ultra-X EDS detector (solid angle >4.45 srad, 5 eV energy dispersion) at 5 μs per pixel averaged over 100 frames.

**Raman spectroscopy**

The Raman spectra in the range of 80 to 2000 cm$^{-1}$ were acquired using a micro-Raman spectrometer (inVia, Renishaw) with a 785 nm diode pumped solid-state laser at room temperature.

**X-ray photoelectron spectroscopy (XPS)**

XPS measurements were performed by a Thermo Fisher Scientific K-Alpha x-ray photoelectron spectrometer using Al Ka radiation (1486.68 eV) under a vacuum degree of $5\times10^{-10}$ Pa. The data was accumulated from 10 cycles of the measurements operating at 15 kV and 10 mA. The pass energy of the measurement was 30 eV, and the step size was 0.05 eV. The spectra were calibrated by referencing the binding energy of carbon (C 1s, 284.8 eV).

**High-energy synchrotron XRD**

High-energy synchrotron XRD (HEXRD) measurements were performed at the P02.1 beamline at PETRA III, Deutsches Elektronen-Synchrotron (DESY) in Hamburg, Germany. Before the measurements, all the powder samples were densely filled into Kapon capillaries (Cole-Parmer Instrument Company, Polyimide tubing 1.132 mm) with an inner diameter of 1.03 mm and a 0.051 mm wall thickness to achieve good transmission of X-ray. All the measurements were performed at room temperature using an exposure time of 600 s. The measured data were acquired on an amorphous silicon two-dimensional flat panel Varex XRD 4343CT (2880×2880 pixel matrix for 150×150 μm² pixel size) in corner configuration with a quarter ring $Q_{max}$ of 28.1 Å$^{-1}$. The beam spot was 1×1 mm² and the used wavelength λ was 0.207 Å (~60 keV). The diffraction patterns collected by Synchrotron radiation XRD were converted into $Q$-Intensity curves using the *FIT2D* software. Here the wavevector $Q = 4\pi \sin(2\theta)/\lambda$, for the scattering angle $2\theta$ and the X-ray wavelength $\lambda$. The resulting data were further corrected for polarization, sample absorption, Compton scattering, multiple scattering, and fluorescence. Then, through an in-house code (43), the resulting data were further corrected for polarization and sample attenuation to obtain the standardized intensity $I(Q)$. Finally, the $I(Q)$ were normalized by fitting to the coherent atomic form factors $\langle f^2 \rangle$ and $\langle f \rangle^2$ at a high $Q$ value to obtain the total structure factors $S(Q)$ as,

$$S(Q) - 1 = \frac{\beta I(Q) - (\langle f^2 \rangle + I_m + I_c + I_f)}{\langle f \rangle^2} \quad (1)$$

Where $\beta$ is the normalization factor, and $I_m$, $I_c$, and $I_f$ are multiple scattering, Compton scattering, and fluorescence intensity, respectively. The obtained $S(Q)$ covered the $Q$ range of 0.5~16 Å$^{-1}$, which was used to derive the reduced pair distribution function (PDF) $G(r)$ in real space by inverse Fourier transformation with Lorch function $L(Q)$ as (43),

$$G(r) = \frac{2}{\pi} \int_0^{16} Q(S(Q) - 1) \sin(Qr) L(Q) dQ \quad (2)$$

**Liquid $^1$H NMR spectroscopy**

Solution $^1$H NMR spectra of digested samples (in a mixture of DCl (35%)/D$_2$O (0.1 mL) and dimethyl sulfoxide (DMSO)-d$_6$ (0.5 mL)) of desolvated ZIF-62 crystals and glasses (about 10 mg) were recorded on a Bruker Avance III 600 MHz spectrometer at 308 K. Chemical shifts were referenced to the residual portion-solvent signals of DMSO-d$_6$. The spectra were processed with the MestreNova Suite.

**Solid-state MAS NMR spectroscopy**

$^{67}$Zn NMR spectra were acquired at a magnetic field of 20.0 T on a Bruker Avance NEO spectrometer, and a Low-E HX 3.2 mm MAS probe built and designed at the National High Magnetic Field Laboratory (NHMFL). The spin-echo spectra were acquired with magic-angle spinning at 16 kHz, and pulses of 4.5 and 9.0 us at a rf field of 18.5 kHz. A ~2 ms WURST pulse with a sweep range of 16 kHz and rf field of ~18 kHz was applied prior to the spin-echo for signal enhancement. Approximately 80000 to 85000 scans were averaged and Fourier transformed to obtain the spectra for the crystals, while for glasses approximately 90000 to 150000 scans were averaged (44).

$^{15}$N and $^{13}$C CPMAS NMR spectroscopy of ZIF-62, HPG, HPG, and HPG$_{673-30}$ were conducted at the National High Magnetic Field Laboratory (NHMFL) using 3.2 mm MAS probes. The $^{15}$N CPMAS spectra were acquired at 10 kHz MAS, with 71.4 kHz 1H decoupling, contact time of 10 ms, recycle

delay of 1 s, and a total of 65536 scans. The $^{13}$C CPMAS spectra were acquired at 10 kHz MAS, with 71.4 kHz 1H decoupling, contact time of 1 ms, recycle delay of 1 s, and a total of 6144 scans.

**Absorption spectroscopy**

Absorption spectra in a region from 200 to 800 nm of all samples were collected by a double beam Agilent's Cary 5000 UV/VIS/IR spectrophotometer. The optical transmittance was obtained by using the equation derived from the Beer-Lambert Law (45).

**Photoluminescence spectroscopy**

The two-dimensional (excitation-emission) photoluminescence (PL) spectra in the region of 260-800 nm were measured by using a fluorescence spectrophotometer (Edinburgh Instruments, FLS1000) equipped with a 450 W Xenon lamp as the excitation source and a PMT980 photomultiplier tube as detector. The PL decay curves were recorded on Edinburgh Instruments FLS980 equipped with both continuous (450 W) and microsecond pulsed xenon lamps, and a Hamamatsu R2658P photomultiplier tube as detector. For temperature-dependent PL spectra at 10-298 K, the samples were mounted on an optical cryostat (DE202, Advanced Research Systems). For the temperature-dependent PL spectra and lifetime at 298-473 K, the samples were mounted on a thermal stage (77-873 K, THMS 600, Linkam Scientific Instruments). The corresponding lifetime curves can be well fitted by a third-order exponential decay mode equation as,

$$I = A_1 e^{(-t/\tau_1)} + A_2 e^{(-t/\tau_2)} + A_3 e^{(-t/\tau_3)} \qquad (3)$$

where $I$ is the PL intensity, $A_1$, $A_2$, and $A_3$ are the fitting parameters, $t$ is the time, $\tau_1$, $\tau_2$, and $\tau_3$ are the decay components. Based on Eq. (3), the average lifetime $\tau$ can be calculated as,

$$\tau = (A_1 \tau_1^2 + A_2 \tau_2^2 + A_3 \tau_3^2)/(A_1 \tau_1 + A_2 \tau_2 + A_3 \tau_3) \qquad (4)$$

The PL quantum yield (PLQY) was measured by using an optical integrating sphere (N-M01, Edinburgh Instruments). The optical response of the instrument was calibrated with a standard tungsten halogen lamp. Subsequently, the samples were put in a cuvette with an optical length of 1 cm. The absolute internal PLQY, the absorbance yield, and the external PLQY can be calculated by using the equations as,

$$\eta_{int} = \frac{N_1}{N_2} \times 100\% \qquad (5)$$

$$\eta_{abs} = \frac{N_2}{N_3} \times 100\% \qquad (6)$$

$$\eta_{ext} = \frac{N_1}{N_3} \times 100\% \qquad (7)$$

where $\eta_{int}$ is internal PLQY, $\eta_{int}$ is the absorbance yield, $\eta_{ext}$ is external PLQY, $N_1$ is the number of emitted photons, $N_2$ is the number of absorbed photons, and $N_3$ is the number of emitted photons by the lamp. Statistical analysis was performed through the Origin software.

**LEDs Fabrication**

LED-1 was packed by integrating a 365 nm UV chip and HPG (2.2 mm thickness); LED-2 was prepared by integrating a 365 nm UV chip (as the excitaor) and HPG$_{673-30}$ (2 mm thickness); LED-3 was prepared by integrating a 365 nm UV chip, HPG$_{673-30}$ (2 mm thickness), and the commercial red phosphor (CaAlSiN$_3$: Eu$^{2+}$). LED-4 was prepared by integrating a 400 nm blue chip and HPG (2.2 mm thickness); LED-5 was prepared by integrating a 400 nm blue chip and HPG$_{673-30}$ (2 mm thickness); LED-6 was prepared by integrating a 400 nm blue chip, HPG$_{673-30}$ (2 mm thickness), and the commercial red phosphor (CaAlSiN$_3$: Eu$^{2+}$). The chip and bulk glass for each LED were glued by using silicone, which was mixed with part A and part B (SHIN-ETSU Chemical Co. Ltd). The obtained LED devices were further cured at 393 K for 2 h in an oven, and then fixed in a 5054 PCB substrate to produce the desired LEDs. The photoelectric properties, including PL spectra, CCT, CRI, and chromaticity coordinate of the fabricated LEDs were measured by an integrating sphere spectroradiometer system (HASS-2000, 350-1100 nm, Everfine). The temperature of the operating

LED under various driving currents was measured by an infrared thermal imager system (E75, FLIR Systems).

**Density functional theory (DFT) calculations**

The electronic structure of ZIF-62 crystal was calculated via using the CP2K package (46). Herein, the Perdew-Burke-Ernzerhof (PBE) functional was employed to describe the system (47). Structural optimization was carried out with convergence criteria including a maximum geometry change of $3 \times 10^{-3}$ Å, an RMS geometry change of $1.5 \times 10^{-3}$ Å, a maximum force of $4.5 \times 10^{-4}$ (arbitrary unit), and an RMS force of $3 \times 10^{-4}$ (arbitrary unit). To account for dispersion interactions, the DFT-D3 correction with Becke–Johnson damping was employed (48, 49). Based on the optimized structure, unrestricted Kohn-Sham DFT was used as the electronic structure calculation method. The Goedecker-Teter-Hutter (GTH) pseudopotentials, DZVP-MOLOPT-SR-GTH basis sets were used to describe the molecules (50). A plane-wave energy cut-off of 800 Ry was employed.

# Acknowledgments

We thank Tianzhao Xu for help with ZIF-62 crystal synthesis and Wessel M. W. Winter and Sabyasachi Sen for the discussion on the results. Part of this work was performed at the National High Magnetic Field Laboratory (NHMFL) in Florida, USA, which is funded by the U.S. National Science Foundation (DMR-2128556) and the State of Florida. We acknowledge the Carlsberg Foundation (Grant no. CF21-0371) for funding the STA 449 F3 ASC used in this work and the Villum Fonden (Grant NO. 58844) for conducting TEM experiments. We also acknowledge the P02.1 beamline (proposal no. I-20240507) at PETRA III, Deutsches Elektronen-Synchrotron (DESY) in Hamburg, Germany.

# Author contributions

Z.C.L. and Y.Z.Y. conceived the project. Z.C.L. synthesized and characterized all samples using DSC, XRD, NMR, and Raman spectroscopy. Z.H.W. performed temperature-dependent PL spectra and lifetime spectra of the samples and fabricated the LED devices. F.M.C. performed high energy XRD experiments, X.G. and M.M.S. provided the beam time, and analyzed the PDF data. H.T.Z. performed 2D PL spectra, lifetime, and absolute internal PLQY measurements of the studied samples. P.B. performed TEM characterization, P.B. and J.J. analyzed the TEM data, and revised the TEM interpretation. L.R.J. analyzed the Raman spectroscopic data. I.H. and Z.H.G. performed NMR measurements and analyzed the data. B.Z.Y. and G.P.D. performed and analyzed the DFT calculations of ZIF-62 crystal. L.C. sintered the HPG for further heat treatment. J.B.Q., F.G., D.H.Y., and H.M.Z. detailly discussed the results. Z.C.L. wrote the manuscript and Y.Z.Y. revised the manuscript. All authors participated in discussing the data and revising the manuscript.

## Competing interests

The authors declare no competing interests.